\documentclass[prl,twocolumn,10pt,showpacs,superscriptaddress,longbibliography,footnoteinbib]{revtex4-1}
\usepackage{amsfonts,amsmath,amssymb,graphicx}

\usepackage{color,times}
\usepackage{soul}

\definecolor{darkblue}{rgb}{0, 0, 0.8}
\usepackage[colorlinks=true, breaklinks=true, linkcolor=darkblue, citecolor=darkblue, urlcolor=darkblue]{hyperref}

\begin{document}

\title{Polaron-Polaritons in the Integer and Fractional Quantum Hall Regimes}

\author{Sylvain Ravets}
\affiliation{Institute of Quantum Electroncis, ETH Z\"urich,\\
CH-8093 Z\"urich, Switzerland}

\author{Patrick Kn\"uppel}
\affiliation{Institute of Quantum Electroncis, ETH Z\"urich,\\
CH-8093 Z\"urich, Switzerland}

\author{Stefan Faelt}
\affiliation{Institute of Quantum Electroncis, ETH Z\"urich,\\
CH-8093 Z\"urich, Switzerland}
\affiliation{Solid State Physics Laboratory, ETH Z\"urich,\\
CH-8093 Z\"urich, Switzerland}

\author{Ovidiu Cotlet}
\affiliation{Institute of Quantum Electroncis, ETH Z\"urich,\\
CH-8093 Z\"urich, Switzerland}

\author{Martin Kroner}
\affiliation{Institute of Quantum Electroncis, ETH Z\"urich,\\
CH-8093 Z\"urich, Switzerland}

\author{Werner Wegscheider}
\affiliation{Solid State Physics Laboratory, ETH Z\"urich,\\
CH-8093 Z\"urich, Switzerland}

\author{Atac Imamoglu}
\affiliation{Institute of Quantum Electroncis, ETH Z\"urich,\\
CH-8093 Z\"urich, Switzerland}

\begin{abstract}
Elementary quasi-particles in a two dimensional electron system can be described as exciton-polarons since electron-exciton interactions ensures dressing of excitons by Fermi-sea electron-hole pair excitations. A relevant open question is the modification of this description when the electrons occupy flat-bands and electron-electron interactions become prominent. Here, we perform cavity spectroscopy of a two dimensional electron system in the strong-coupling regime where polariton resonances carry signatures of strongly correlated quantum Hall phases. By measuring the evolution of the polariton splitting under an external magnetic field, we demonstrate the modification of electron-exciton interactions that we associate with phase space filling at integer filling factors and polaron dressing at fractional filling factors. The observed non-linear behavior shows great promise for enhancing polariton-polariton interactions.

\end{abstract}


\maketitle

Strong coupling of excitons in a semiconductor quantum well (QW) to a microcavity mode leads to formation of quasiparticles called cavity exciton polaritons~\cite{Weisbuch1992}. Polaritons have played a central role in the investigation of nonequilibrium condensation and superfluidity of photonic excitations~\cite{Deng2010,Carusotto2013}. While polaritons acquire a finite nonlinearity due to their exciton character, interactions between polaritons in undoped QWs are not strong enough for realizing strongly interacting photonic systems~\cite{Amo2016}.

Two-dimensional electron systems (2DES) evolving in large magnetic fields, in contrast, are a fertile ground for many-body physics due to prominence of electron-electron interactions. Formation of skyrmion excitations in the vicinity of filling factor $\nu = 1$ is a consequence of such interactions. More spectacularly, electron correlations lead to the formation of fractional quantum Hall (FQH) states where the ground state exhibits topological order~\cite{Prange1989,DasSarma1997,Sondhi1997}. Moreover, it has been proposed that a sub-class of FQH states exhibit non-abelian quasi-particles which can be used to implement topological quantum computation~\cite{Nayak2008}. The nature of optical excitations of a 2DES have also recently generated lots of interest: experimental and theoretical~\cite{Sidler2017,Efimkin2016} studies in transition metal dichalcogenide (TMD) monolayers have established that these excitations should be described in the framework of the Fermi polaron problem, as a collective excitation resulting from exciton-electron interactions~\cite{Schirotzek2009,Koschorreck2012,Schmidt2012,Massignan2014}. In this context, optical excitation of an exciton leads to generation of an electron screening cloud that results in formation of a lower energy attractive exciton-polaron~\cite{Sidler2017,Efimkin2016}. In this work, we report corresponding signatures in GaAs, where energy scales are known to differ significantly compared to TMD monolayers~\cite{Koudinov2014,Suris2003}, due to a particularly small binding energy of the bound-molecular trion state. More importantly, our work presents experimental signatures of polaron formation at nonzero magnetic fields where electrons are confined to the lowest Landau level.

\begin{figure}
\centering
\includegraphics[width=86mm]{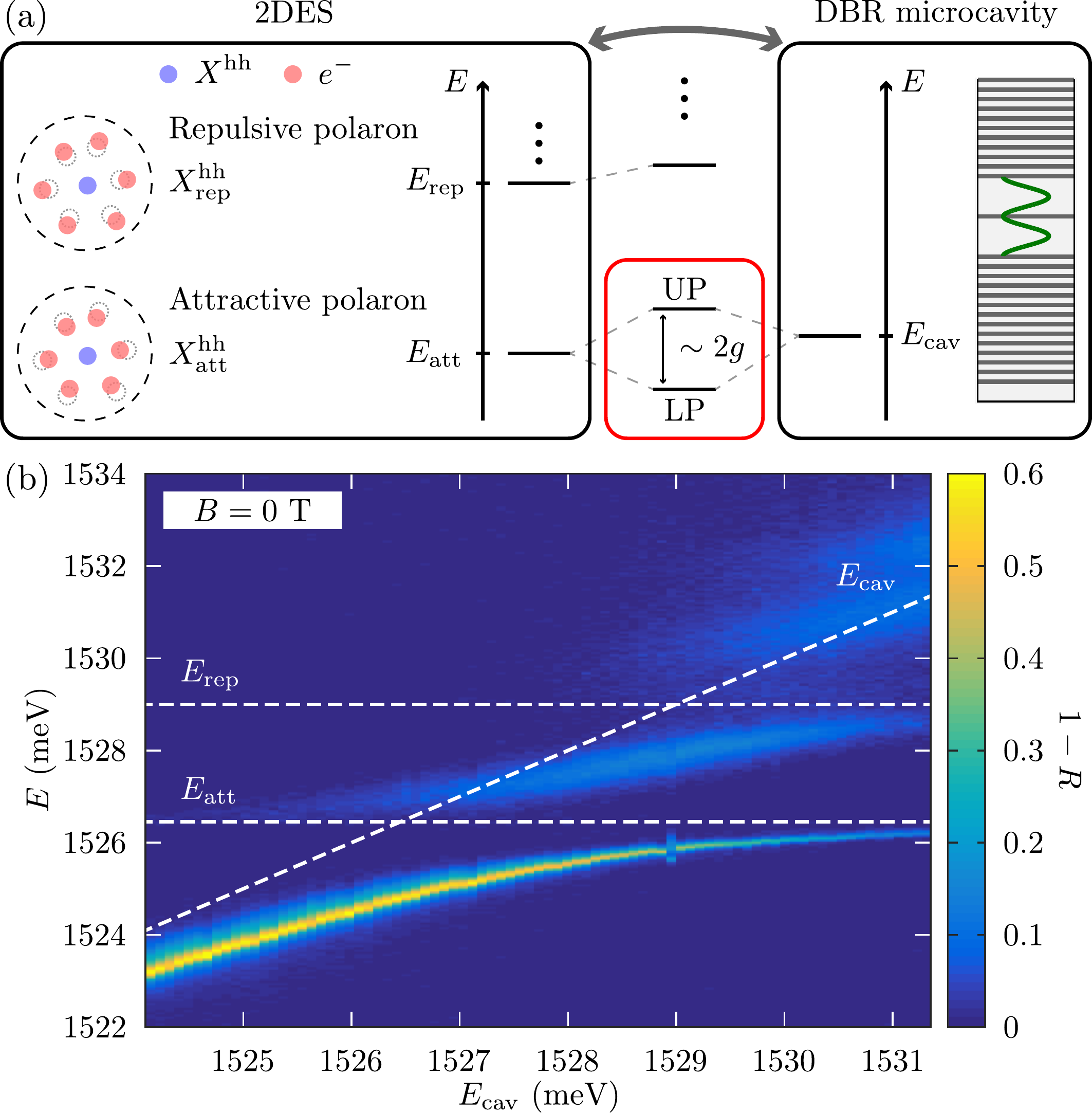}
\caption{Observation of polaron-polaritons in a GaAs QW. (a)~We couple a 2DES to the optical mode of a microcavity composed of two distributed Bragg reflectors (DBRs). In the strong coupling regime, the lowest energy eigenstates of the coupled system are the lower polariton (LP) and the upper polariton (UP). (b)~Reflectivity spectrum of the system as we tune the cavity frequency.}
\label{fig:Spectroscopy}
\end{figure}

\begin{figure*}
\centering
\includegraphics[width=172mm]{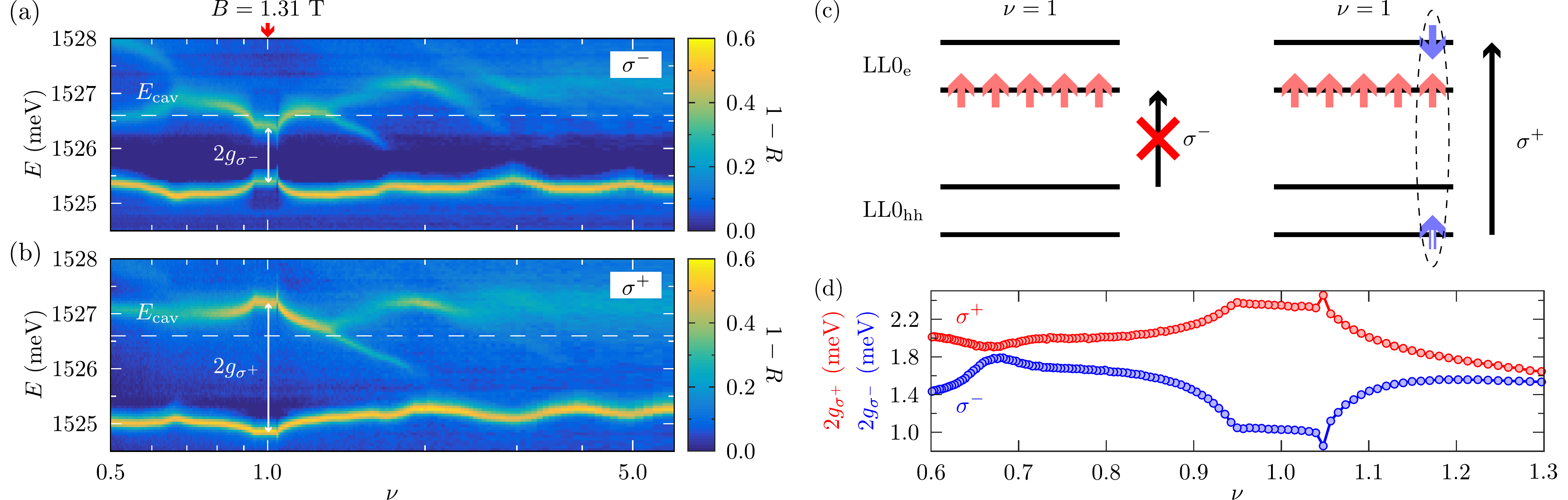}
\caption{Cavity spectroscopy of the system in the fractional quantum Hall regime, as we tune $B$ for fixed $E_{\rm cav}$ (white dashed line). Measurement performed with (a)~$\sigma^-$ and (b)~$\sigma^+$ polarized light. (c)~Relevant energy levels and optical transitions around $\nu=1$. (d)~Polariton splitting measured in $\sigma^-$ (blue) and $\sigma^+$ (red) polarizations.}
\label{fig:IQHE}
\end{figure*}

It has recently been demonstrated that embedding a 2DES inside a microcavity realizes an alternate method for probing quantum Hall (QH) states~\cite{Smolka2014}. In the strong coupling regime, polariton excitations are sensitive to elementary properties of the many-body ground state, such as spin-polarization and incompressibility due to their part-exciton character. In contrast to bare excitons though, polaritons are immune to decoherence processes such as phonon or impurity scattering due to their ultra-light mass, ensuring that they are delocalized. Consequently, the energy resolution achievable in polariton-based spectroscopy is only limited by the polariton decay rate due to mirror losses, which can be on the order of 20~mK in state-of-the-art microcavities~\cite{Steger2013}. In the present work, we also demonstrate a new feature of cavity-polariton-spectroscopy of FQH states: by adjusting the separation distance between the 2DES and the doping layers we substantially reduce unwanted light-induced variations of the 2DES electron density $n_e$~\cite{Supplemental}.

\begin{figure*}
\centering
\includegraphics[width=172mm]{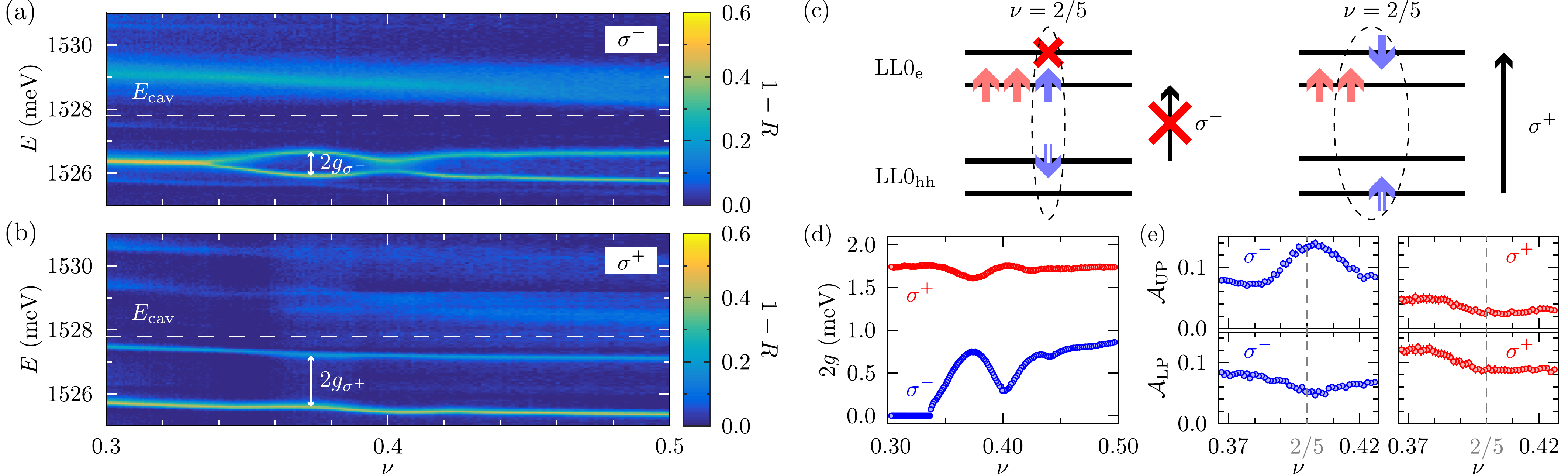}
\caption{Cavity spectroscopy of the system in the fractional quantum Hall regime, as we tune $B$ for fixed $E_{\rm cav}$ (white dashed line). Measurement performed with (a)~$\sigma^-$ and (b)~$\sigma^+$ polarized light. (c)~Relevant energy levels and optical transitions around $\nu=2/5$. (d)~Polariton splitting around $\nu=2/5$ measured in $\sigma^-$ (blue) and $\sigma^+$ (red) polarizations. (e)~Lower polariton ($\mathcal{A}_{\rm LP}$) and upper polariton ($\mathcal{A}_{\rm UP}$) peak areas in $\sigma^-$ (blue) and $\sigma^+$ (red) polarizations (arbitrary units).}
\label{fig:FQHE}
\end{figure*}

Our structure consists of a 2DES in a 20~nm modulation doped GaAs QW, embedded at the center of a $2 \lambda$ $\rm Al_{0.19}Ga_{0.81}As$ microcavity ~\cite{Supplemental}. The front (back) distributed Bragg reflector (DBR) is composed of 19 (25) pairs of $\rm AlAs / Al_{0.20}Ga_{0.80}As$ layers leading to the measured quality factor $Q \simeq (5.5 \pm 0.1) \times 10^3$ for the cavity. The QW features a double-sided silicon $\delta$-doping with a set-back distance of $3\lambda / 4$ above and below the center of the cavity. From magneto-transport measurements~\cite{Koon1992}, we estimate the 2DES electron density $n_e \simeq 0.33 \times 10^{11}~{\rm cm}^{-2}$ and the mobility $\mu \simeq 1.6 \times 10^6~{\rm cm}^2{\rm V}^{-1}{\rm s}^{-1}$. We deliberately choose a relatively low $n_e$ to access the physics of the lowest Landau level in the range of magnetic fields currently available on our experimental setup ($|B| \leq 8~{\rm T}$). The relatively high $\mu$ ensures that we can probe FQH physics.

We perform polarization-resolved spectroscopy of the 2DES using an infrared light emitting diode centered around 820~nm. We shine excitation light onto the sample placed in a dilution refrigerator with a 30~mK base temperature. An aspheric lens (${\rm NA}=0.15$) collects the light reflected off the sample, which is analyzed using a spectrometer. Earlier studies on the optics of 2DES have shown extreme sensitivity of $n_e$ to optical power~\cite{Kukushkin1989,Goldberg1992,Groshaus2007,Smolka2014}. Increasing the optical power not only changes $n_e$, but also causes qualitative changes in the reflectivity spectrum~\cite{Wuster2015}, which is detrimental to the study of fragile QH states. These unwanted effects are attributed to photoexcitation of $DX$ centers in Si-doped ${\rm Al}_x {\rm Ga}_{1-x} {\rm As}$ with $x > 0.2$~\cite{Ihn2010}. We minimize light-induced variations of $n_e$ by keeping $x < 0.2$ in the structure, and more importantly by placing the dopants in 10~nm GaAs ($x=0$) doping quantum wells (DQW). Further, we locate the DQWs in nodes of the electric field inside the cavity~\cite{Supplemental}, which minimizes the intracavity light intensity at the position of the dopants.

We first carry out cavity spectroscopy of the 2DES (see Fig.~\ref{fig:Spectroscopy}a) when $B=0~{\rm T}$. Fig.~\ref{fig:Spectroscopy}(b) shows the reflectivity spectrum as a function of $E_{\rm cav}$. In contrast to undoped QW structures~\cite{Weisbuch1992}, we observe coupling to three exciton-like resonances as we scan $E_{\rm cav}$. For the lowest energy anticrossing, we measure a normal mode splitting of $2g=2.00 \pm 0.01~{\rm meV}$. Since $2g$ is larger than the bare-cavity linewidth $\gamma_c \simeq 280 \pm 10~{\rm \mu eV}$, the system is in the strong coupling regime of cavity-QED and the elementary excitations should be characterized as cavity-polaritons~\cite{Weisbuch1992}. Since the cavity-exciton coupling in this system is comparable to energy level splittings of the three exciton-like resonances, the polariton modes observed in the reflection spectrum can only be described as a superposition of all underlying resonances (see Fig.~\ref{fig:Spectroscopy}a). We identify the lowest energy exciton-like resonance observed in Fig.~\ref{fig:Spectroscopy}(b) as the heavy-hole attractive polaron ($X^{\rm hh}_{\rm att}$) -- a heavy-hole exciton dressed by Fermi sea electron-hole pair excitations~\cite{Sidler2017}. Since the attractive polaron resonance is associated with the bound-molecular singlet trion channel, it was previously referred to as ``trion mode''~\cite{Esser2001,BarJoseph2005}. We assign the middle-energy excitonic resonance to the heavy hole repulsive-polaron ($X^{\rm hh}_{\rm rep}$)~\cite{Sidler2017,Efimkin2017}. The magnitude of the splitting of this mode from the attractive polaron ($2.5$~meV) is a factor of 2 larger than the bare trion binding energy and is fully consistent with its identification as the repulsive polaron branch. Finally, we tentatively identify the highest energy excitonic mode to the light-hole exciton~\footnote{The absence of attractive and repulsive polaron branches associated with the light-hole exciton may stem from the absence of a bound light-hole trion, or from the lower oscillator strength and the larger broadening of these resonances}.

Next, we analyze the $B \neq 0$ case where the electrons are confined to the lowest Landau Level (LL), with filling factor $\nu$~\cite{Prange1989}. To explore the interplay between quantum Hall states and polaritonic excitations, we tune $E_{\rm cav}$ to ensure that the cavity mode dressed by nonperturbative coupling to higher energy excitonic modes is resonant with $X^{\rm hh}_{\rm att}$. Since the lowest energy polariton has predominantly $X^{\rm hh}_{\rm att}$ character, the spin state of the optically generated electron is determined by the photon polarization~\cite{Smolka2014}: left-hand circularly polarized light $\sigma^-$ probes transitions to the lower electron Zeeman spin subband ($\left | \uparrow \right >$) and right-hand circularly polarized light $\sigma^+$ probes transitions to the upper electron Zeeman spin subband ($\left | \downarrow \right >$). Consequently, the observed spectral signatures are strongly dependent on how the electrons are arranged in the LLs i.e.\ on the spin-polarization of the different ground states of the 2DES~\cite{Goldberg1992,Groshaus2007}.

Figure~\ref{fig:IQHE}(a-b) shows the white light reflection spectrum as a function of $\nu$, varied by scanning $B$. Here, we tuned $E_{\rm cav}$ close to resonance with the $\left | \uparrow \right >$-transition of lowest Landau level LL0 at $\nu=1$. The most striking feature is the collapse of $g_{\sigma^-}$ around $B=1.31~{\rm T}$, concurrent with the enhancement of $g_{\sigma^+}$. We associate this feature with the $\nu=1$ QH state, in excellent agreement with the value of the electron density measured independently. The observed behavior is a direct consequence of the high degree of spin-polarization of the QH ferromagnet at $\nu = 1$: for a fully polarized state, $g_{\sigma^-}$ is expected to collapse due to the fact that all $\left | \uparrow \right >$-electron states are occupied. Phase space filling thus prevents optical excitation of an electron to that level, and therefore the oscillator strength for that transition collapses (see Fig.~\ref{fig:IQHE}(c)). Concurrently, all $\left | \downarrow \right >$-electron states are free and $g_{\sigma^+}$ increases due to the increased number of available states~\cite{Aifer1996,Groshaus2004}. Fig.~\ref{fig:IQHE}(d) shows $g_{\sigma^-}$ and $g_{\sigma^+}$ extracted from fits of the reflection spectra. From this, we calculate the spin-polarization $S_z \simeq ( g_{\sigma^+}^2 - g_{\sigma^-}^2 ) / ( g_{\sigma^+}^2 + g_{\sigma^-}^2 )$ at $\nu=1$~\cite{Groshaus2007,Smolka2014}. We obtain $S_z \simeq 70~\%$~\cite{Supplemental}, suggesting that full polarization is not achieved at $\nu=1$, contrary to what is expected for the quantum Hall ferromagnet. In our low $n_e$ sample, incomplete polarization may arise due to disorder and reduced screening of impurity potentials~\cite{Manfra1997}. Furthermore the cyclotron frequency is comparable to the exciton binding energy, ensuring that exciton formation has a sizable contribution from higher LLs. As a consequence, our measurements only yield a lower-bound on $S_z$, and we do not expect full cancellation of $g_{\sigma^-}$ at $\nu=1$. We finally observe a rapid, symmetric depolarization on both sides of $\nu=1$ which is compatible with formation of many-body spin excitations in the ground state (skyrmions and anti-skyrmions)~\cite{Aifer1996,Manfra1997,Zhitomirsky2002,Groshaus2004,Plochocka2009} as a consequence of the competition between Coulomb and Zeeman energies. Finally, coupling to $\nu > 1$ integer QH states is also visible in Fig.~\ref{fig:IQHE}(a-b) as variations of the lower polariton energies vs $\nu$ as a consequence of phase space-filling~\cite{Supplemental}. We emphasize that these spectral features are robust against increased optical powers, which demonstrates that our sample structure provides, through ``cavity protection'' of the 2DES, a unique platform for optical studies of QH physics~\cite{Supplemental}.

We investigate FQH states by scanning $B$ to up to 5~T for an increased value of $E_{\rm cav}$ as shown in Fig.~\ref{fig:FQHE}(a-b). Increasing $B$ reduces $\nu$, thus leading to absorption in a partially filled lowest LL~\cite{Goldberg1990,Yusa2001,Byszewski2006}. Cavity coupling to several FQH states is observed in Fig.~\ref{fig:IQHE} and Fig.~\ref{fig:FQHE} as a $\nu$-dependent normal mode splitting in both polarizations. Such spectral signatures are particularly striking when $\nu$ reaches the fractional values $\nu = 1/3$, 2/5, 2/3 and 5/3. We observe that $g_{\sigma^-}$ and $g_{\sigma^+}$ differ significantly at $\nu = 1/3$, 2/5 and 5/3, which shows that these fractional QH states experience sizable spin-polarization~\cite{Supplemental}. On the contrary, $g_{\sigma^-} \simeq g_{\sigma^+}$ at $\nu=2/3$ shows that this state is not polarized, as expected for samples with $n_e$ in the range of the one studied here. Increasing $n_e$ should allow us to probe the phase transition from an unpolarized to a polarized $2/3$-state~\cite{Eisenstein1990,Smolka2014}.

We now focus on filling factor $\nu=2/5$, see Fig.~\ref{fig:FQHE}(a-b). In stark contrast with the integer QH states, both $\left | \uparrow \right >$ and $\left | \downarrow \right >$ states are available and phase-space filling only plays a marginal role here. Fig.~\ref{fig:FQHE}(d) shows polariton splittings $g_{\sigma^-}$ and $g_{\sigma^+}$ extracted from fits of the reflectivity spectra. One striking feature is that the collapse of $g_{\sigma^-}$ around $\nu=2/5$ is not accompanied with an appreciable increase in $g_{\sigma^+}$, contrary to what was observed for $\nu=1$. Because the LLs are partially filled, the mechanism leading to modification of the polariton splitting is indeed modified.

We argue that the decrease $g_{\sigma^-}$ for a spin-polarized state is due to the polaron nature of optical excitations that are accessible when promoting an electron into the $\left | \uparrow \right >$-state with $\sigma^-$-polarized light. For a fully polarized state, all electrons are in the same $\left | \uparrow \right >$-state and there are no electrons in the $\left | \downarrow \right >$-state. Since the oscillator strength of the $\sigma^-$ singlet $X^{\rm hh}_{\rm att}$ is proportional to the density of $\left | \downarrow \right >$ electrons, perfect spin polarization would lead to vanishing cavity coupling. In contrast, promotion of an electron in the $\left | \downarrow \right >$-state with $\sigma^+$-polarized light always leads to formation of a singlet polaron excitation with electrons available in the $\left | \uparrow \right >$-state, and the polariton splitting is only marginally modified. Fig.~\ref{fig:FQHE}(e) plots the evolution of the polariton peak areas around $\nu=2/5$. The decrease in polariton splitting in $\sigma^-$-polarization is accompanied with a loss (gain) of weight of the lower (upper) polariton. This observation is fully consistent with a reduced cavity-polaron coupling strength and a finite detuning between the bare polaron and cavity resonances, ensuring that the lower (upper) polariton has predominantly polaron (cavity) character at $\nu=2/5$. The absence of a similar oscillator strength transfer in $\sigma^+$-polarization on resonance further supports the interpretation of our data in terms of inhibition of $\sigma^-$ polaron-dressing by $\left | \downarrow \right >$-electrons at $\nu=2/5$.

\begin{figure}
\centering
\includegraphics[width=86mm]{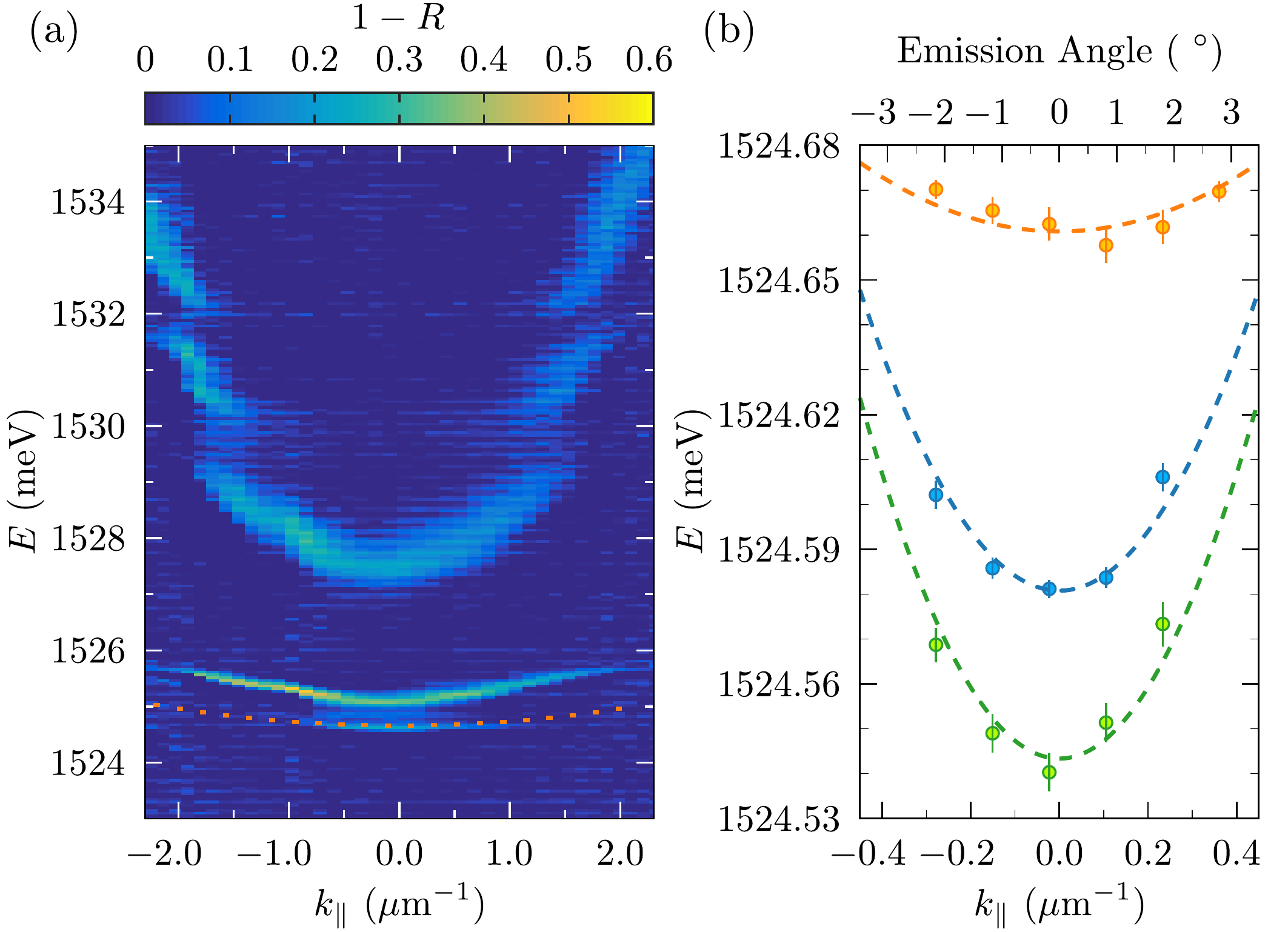}
\caption{Polaron-polariton dispersion around $\nu=2/5$. (a)~Cavity spectroscopy exactly at $\nu=0.4$ for different in-plane momenta $k_{\parallel}$. The flat reflection signal observable between the lower and upper polaritons around $k=0~{\rm \mu m}^{-1}$ is an experimental artefact, stemming from an etalon effect in the detection path. (b)~Energy of the lower polaron-polariton for small $k_{\parallel}$ at filling factor $\nu=0.42$ (green), $\nu=0.4$ (orange) and $\nu=0.37$ (blue). Dashed lines show parabolic fits to the lower polariton energies.}
\label{fig:Dispersion}
\end{figure}

Finally, we address the question of the modification of the polaron-polariton effective mass in the vicinity of $\nu=2/5$. We use a $\mathrm{NA}= 0.68$ lens to excite a broad range of in-plane momenta $k_{\parallel}$ using the same broadband light emitting diode. A low NA lens couples the reflected light into a fiber, which enables angle selective measurements. The dispersion relation in Fig.~\ref{fig:Dispersion}(a) at $\nu=2/5$ clearly shows the anticrossings with $X^{\rm hh}_{\rm att}$ and $X^{\rm hh}_{\rm rep}$ as pointed out already in Fig.~\ref{fig:Spectroscopy}(b). We fit a parabola to the lower polariton dispersion at $\nu=2/5$ (dashed orange line) and compare it, in Fig.~\ref{fig:Dispersion}(b), to the dispersions measured at filling factors slightly above (green) and below (blue). Strikingly, we find an increase of the effective mass $m^*$ at $\nu=2/5$ (orange) by a factor of $\simeq 4 \pm 2$ compared to $\nu=0.42$ (green) and $\nu=0.37$ (blue)~\footnote{Note that estimation of the lower polariton mass at $\nu = 2/5$ is rendered difficult due to the low curvature of the parabola.}. This observation illustrates further the strong reduction in the oscillator strength of the attractive-polaron resonance which reduces the cavity-character and enhances $m^*$.

We emphasize that theory of exciton-polarons has been previously developed for excitons interacting with a 2DES in the limit $B=0$~\cite{Sidler2017,Efimkin2016}. A quantitative modeling of our experiment requires extending prior theoretical work to the case of screening of excitons by electrons occupying a single LL: a significant advance in this direction was the recent development of the theory of exciton-polarons in the limit of strong magnetic fields but without taking into account electron-electron interactions leading to FQH states~\cite{Efimkin2017}. Our work focused on the singlet channel which plays a prominent role in the limit of moderate magnetic fields ($B \le 3.5$~T) used in our experiments. Yet, we expect triplet channels to play a key role in determining the full polariton spectrum, particularly at higher B-fields relevant for samples with higher electron density. A more challenging problem is exciton-electron interactions in the vicinity of FQH states: polaron-polariton formation in this limit may be described using polariton dressing by fractionally charged quasi-particle-hole pairs~\cite{Grusdt2016}. The latter problem is related to identification of signatures of incompressibility of the many-body ground state in the polariton excitation spectrum.

On the technical side, we demonstrate that cavity electrodynamics is an invaluable platform to probe fragile fractional states. This could potentially enable optical manipulation of anyonic quasi-particles associated with strongly-correlated phases. Furthermore, increasing the quality factor of the cavity could further enhance the sensitivity of our measurements~\cite{Steger2013}. Finally, the observed filling factor-dependent polariton splitting could be particularly useful to engineer single-photon non-linearities between polaritons and allow for experimental realization of strongly correlated driven-dissipative photonic systems in arrays of semiconductor microcavities~\cite{Amo2016}.

\begin{acknowledgments}
The Authors acknowledge many useful discussions with Hadis Abbaspour, Valentin Goblot, Wolf Wuester and Sina Zeytinoglu. This work was supported by NCCR Quantum Photonics (NCCR QP), an ETH Fellowship (S. R.), and an ERC Advanced investigator grant (POLTDES).
\end{acknowledgments}


%

\end{document}